# SHELL MODEL APPLICATIONS TO $N = Z$ NUCLEI[*]

Gabriel Martínez-Pinedo

Department für Physik und Astronomie, Universität Basel, CH-4056 Basel, Switzerland

This manuscript reviews recent results of large scale shell model calculations on $N = Z$ nuclei. The topics discussed include the role proton-neutron pairing on the binding energies of $pf$-shell nuclei and his influence on backbending rotors; coulomb energy differences and their relationship with alignment. Proton-neutron correlations are found responsible for the characteristic features of the isotopic shifts in calcium, the parabolic dependence on $A$ and the prominent odd-even staggering.

## 1. Introduction

In the recent years, thanks to the enormous progress done in computer capabilities and programming techniques, shell-model calculations, based on microscopic effective interactions and in model spaces that include all the basic physics ingredients, have become available. The spherical shell model approach has provided an excellent description of many properties, from level schemes to rare processes as double beta decays, and has given a unified description of the single particle and collective degrees of freedom of the nucleus. Taking advantage of the availability of detailed microscopic calculations we will study the role of proton-neutron correlations in $N = Z$ nuclei. We will discuss the effects of pn pairing on the binding energies of $pf$-shell nuclei and his influence on the rotational properties taking $^{48}$Cr as an example. Next we will examine how the alignment of particles evolves as a function of angular momentum and his relationship with the observed coulomb energy differences in mirror nuclei. Finally, we will show how cross shell proton-neutron correlations are responsible for the isotope shifts in calcium.

---

## 2. Pairing and $N = Z$ nuclei

The study of the isovector pairing interaction among like particles is one of the classical themes of nuclear physics. Proton-neutron pairing has been much less studied, in particular its isoscalar part. The advent of high-spin physics and unproven spectroscopic data in heavy $N \approx Z$ nuclei has again brought an old question: how does pairing evolve with angular momentum? We shall examine these issues in what follows. To start we select as our $T = 0$ and $T = 1$ pairing Hamiltonians those extracted in [1] from the realistic G-matrices. We keep their notation and call $P01$ and $P10$ the isovector and isoscalar $L = 0$ pairing Hamiltonians. In the absence of any other interaction their spectra are given by:

$$E_{P01} = -G \left[ \frac{(n - v_s)(4\Omega + 6 - n - v_s)}{8} + \frac{t(t+1)}{2} - \frac{T(T+1)}{2} \right] \quad (1)$$

$$E_{P10} = -G \left[ \frac{(n - v_t)(4\Omega + 6 - n - v_t)}{8} + \frac{s(s+1)}{2} - \frac{S(S+1)}{2} \right] \quad (2)$$

where $G$ is a coupling constant, $\Omega$ is the maximum number of $L = 0$ states in the valence space, $n$ the number of valence particles, $v_s$ and $v_t$ the singlet and triplet seniorities, $t$ and $T$ the reduced and total isospin, and $s$ and $S$ the reduced and total spin. It is interesting to note that while the isovector pairing favors states with good isospin the isoscalar pairing favors states with good spin. Due to the presence of a strong spin orbit term in the nuclear force the presence of a $L = 0$ isoscalar condensate in the ground state of $N = Z$ nuclei is very unlikely [2].

In order to determine the role of the pairing operators in the behavior of different nuclei and different physical quantities, we make first a reference calculation using the interaction KB3 [3] and then we subtract from KB3 the isovector or isoscalar pairing Hamiltonians and make the calculations with the new interactions KB3-P01 and KB3-P10. We obtain the effect of each pairing channel by direct comparison with the reference calculation. The value of the coupling constant $G$ is obtained from the numbers in Table 1 of ref. [1]. We use $G = -0.295$ MeV for P01 and $G = -0.459$ MeV for P10.

Figure 1 shows the contributions to the ground state binding energy from the isovector pairing (labeled P01) and from the isoscalar pairing (labeled P10) for several isotopic chains. The figure shows the strong odd-even staggering of the P01 points, mostly suppressed in the P10. The only little surprise is that for A=46 and 50 moving from $T = 1$ to $T = 0$, not only the contribution of P01 decreases, but also the contribution of P10. This may explain why the ground state of $^{46}$V and $^{50}$Mn has $T = 1$ instead of $T = 0$. The monopole part (symmetry energy) part of KB3 will put the centroid of the $T = 0$ states lower than the centroid of the $T = 1$ states by about 1.3 MeV; on the other hand, the total pairing contribution to $T = 1$ is larger by nearly 2 MeV than the contribution to $T = 0$. Therefore, it is the $T = 1$ state that becomes the ground state of the odd-odd $T_z = 0$ nuclei.

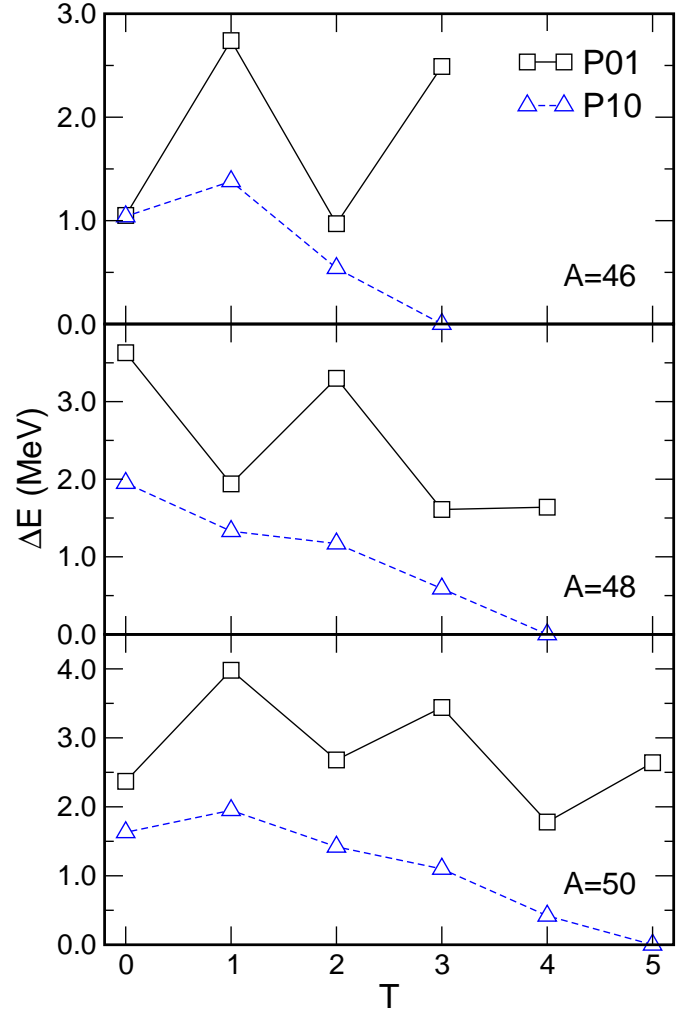

Fig. 1. Pairing contributions to the ground state energies of the $A = 46, 48$, and $50$ isobaric chains as a function of the isospin $(T)$

To study the evolution of the pair content of a nucleus as the rotational frequency increases we have chosen $^{48}$Cr, which is the most representative example. In figure 2a we present as a backbending plot the effect of the pairing correlations, basically a change in the static moment of inertia. For states below the backbending – i.e. for those that can be viewed as proceeding of the same intrinsic, well deformed state – the wavefunctions with or without pairing have the same structure. Their $B(E2)$'s and quadrupole moments are equal within a few per cent and their overlaps are always better than 95%. This means that pairing does not affect the quadrupole properties that have somehow reached saturation in the deformed regime. This suggest that the discrepancies found in the comparison of the CHFB and SM results in ref. [4] are due to deficiencies in the treatment of pairing in the mean field description. To clarify this point we have used the matrix elements of the Gogny force [5] (obtained [6] using the wave functions computed in a spherical Hartree-Fock calculation with the same force). The results are plotted in figure 2b. It is clearly seem that the shell model calculation with the Gogny force now gives a much better moment of inertia than the CHFB calculation with the same force.

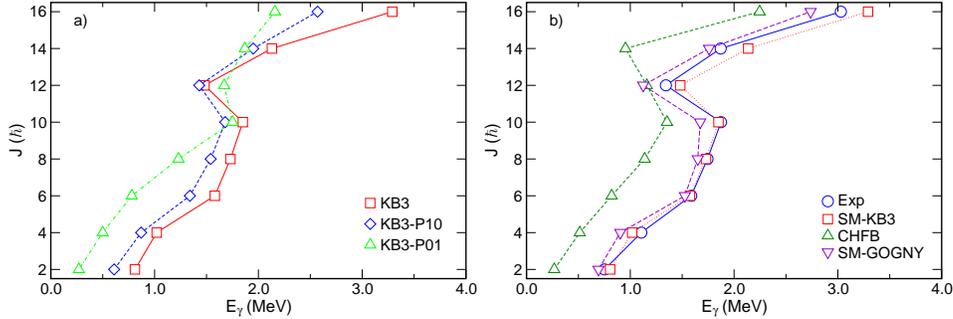

Fig. 2. a) The effect of the isoscalar and isovector pairing in the gamma energies along the yrast line in $^{48}$Cr. b) Gamma energies obtained with the Gogny force in the Shell-Model and cranked Hartree-Fock-Bogoliubov approaches.

## 3. Coulomb displacement energies and alignment

In the previous section we have study the pairing correlations as a functions of the rotational frequency. However, traditionally phenomena as the backbending of the appearance of yrast traps are related to the alignment of particles. In ref. [7] it was show that the $12^+$ yrast trap in $^{52}$Fe can be related to the alignment of two pair of particles. We can use a schematic Hamiltonian to "count" the number of aligned particles:

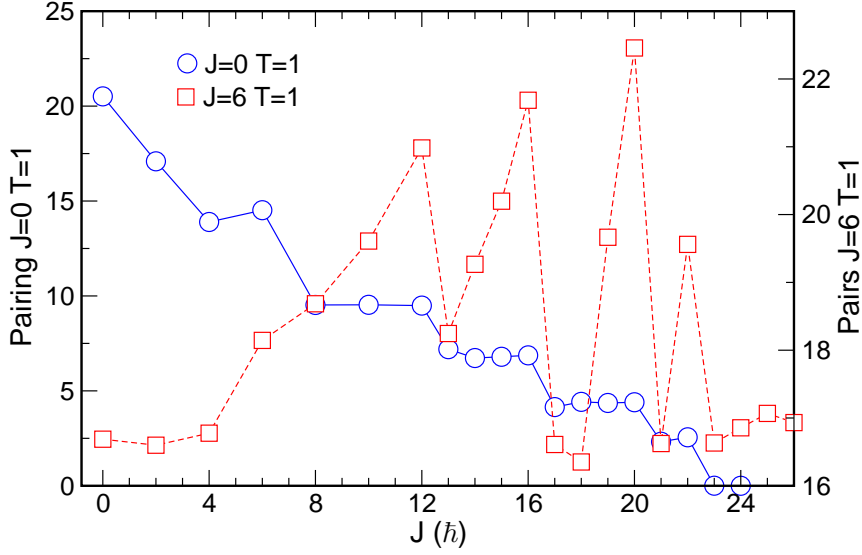

Fig. 3. Expected values of P01 and $H_{\text{align}}$ (arbitrary units) for the yrast states of $^{52}$Fe

$$H_{\text{align}} = \sum_{rs} Z^{\dagger}_{rs}(JT) \cdot Z_{rs}(JT), \qquad (3)$$

where the sum runs over different orbits $(rs)$ and $Z^{\dagger}_{rs}(JT)$ is the creation operator for a normalized pair of particles in the orbits $rs$ with angular momentum $J$ and isospin $T$. As the conmutator $[Z^{\dagger}_{rs}(JT), Z_{rs}(JT)]$ is not a $c$-number the operator above is not a true number operator however it can provide qualitative information about the change in the number of pairs. In figure 3 we present the expected value of $H_{\text{align}}$ ($J = 6$ is the maximum spin with $T = 1$ in the $pf$ shell) for the yrast band of $^{52}$Fe. We include also the expected value of P01 using $G = 1$. For low spin states ($J = 0$–4) $^{52}$Fe presents a rotational spectra. As can be seem in the figure the rotations is collective, no alignment of particles takes place (flat $H_{\text{align}}$). For $J = 6$ a pair of particles is aligned. As $J = 6$ is the maximum possible spin with only a $J = 0$ pair broken a second pair is broken producing a sudden drop in P01. A progressive increase in $H_{\text{align}}$ takes now place and for $J = 12$ two pairs of particles are fully aligned. Note that this sudden alignment of particles could be responsible for the very small $B(E4)$ ($5 \times 10^{-4}$ W.u.) recently measured in a study of the decay of the $12^+$ state [8]. For spins higher than $12^+$ there is a clear correlation between the points of maximum alignment (in the $pf$-shell there could be only 2 pairs with J=6 and T=1)

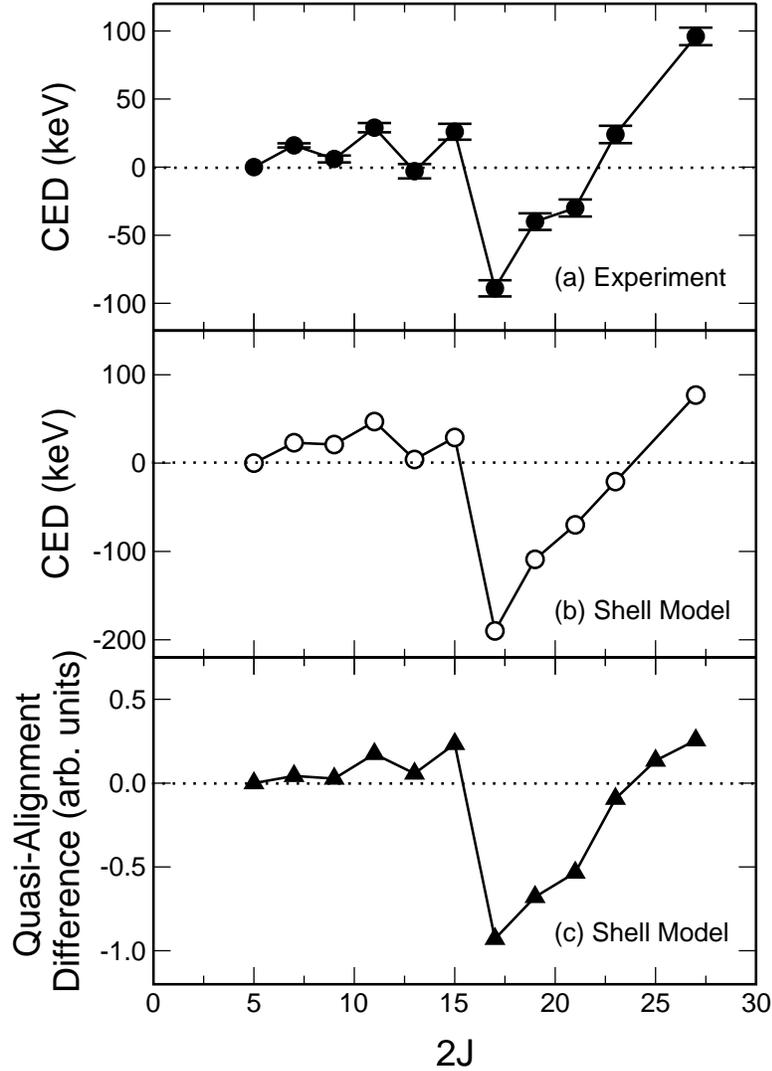

Fig. 4. Experimental (a) and calculated (b) CED defined as $E_x(^{51}\text{Fe} - E_x(^{51}\text{Mn}.$ c) Difference of the proton contribution to $H_{\text{align}}$ in $^{51}$Mn and $^{51}$Fe.

and the breaking of a $J = 0$ pair.

Gamma-ray spectroscopy of high spin states in mirror nuclei has been revitalized in recent years through the development of large gamma-ray spectrometers. Such spectrometers have yielded a wealth of new information on high spin states in $N = Z$ nuclei and proton-rich nuclei in the $f_{7/2}$ shell. The small differences (normally a few tens of keV) observed in the excitations energies are mainly due to the isospin symmetry breaking Coulomb interaction. In recent works [9, 10], the high spin states of $T = 1/2$ mirror nuclei $^{47}$Cr/$^{47}$V and $^{49}$Mn/$^{49}$Cr have been studied. It was found that the CED's are extremely sensitive to the structure of the nuclear wave functions. In recent experiments [11, 12] high spin states up to the band termination in the mirror pair $^{51}$Fe/$^{51}$Mn have been observed. The measured CED's are shown on figure 4 where they are compared with shell-model calculations that include the Coulomb matrix elements denoted "A42" on ref. [9], they were obtained from the $A = 42$ mirror pair. It is seen that there is an abrupt change in the CED's at $J = 17/2^-$. The effect is also present in the calculation, that shows exactly the same trends as the experiment, although with enhanced values. (See ref. [13] for a quantitative reproduction of the measured CED's.) The large increase in the CED can be interpreted as due to the alignment of a pair of protons in $^{51}$Fe which reduces the overlap of their spatial distributions and causes a corresponding reduction in the Coulomb energy. In $^{51}$Mn, the odd proton blocks this alignment and a pair of neutrons is aligned, with no resulting coulomb effect. Beyond $J = 17/2^-$, the protons start aligning also in $^{51}$Mn and therefore the CED's approach zero at the band termination. In order to make more visible we have compute the expected value of the proton contribution to $H_{\text{align}}$, defined in equation (3). Figure 4c shows the difference between the expected values for states of $^{51}$Mn and $^{51}$Fe.

It is interesting to note the similarities between the alignment observed in $^{52}$Fe and the mirror pair $^{51}$Fe/$^{51}$Mn. In fact the aligned $J = 17/2^-$ state in $^{51}$Fe/$^{51}$Mn could be seen as resulting of the coupling of a $j = 7/2^-$ hole to the yrast trap $12^+$ state in $^{52}$Fe. Similarly to what happens in $^{52}$Fe where a very small $B(E4)$ decay probability has been measured, the $17/2^-$ state in $^{51}$Fe/$^{51}$Mn has also very small electromagnetic transitions [14].

## 4. Cross-shell correlations: isotope shifts in Calcium

The appearance of shell gaps associated with magic nucleon numbers is one of the cornerstones of nuclear structure. However, it has become increasingly evident in recent years that these magic numbers, and the corresponding shell closures, might get eroded with increasing neutron excess. A prominent example is the magic neutron number $N = 20$ which vanishes in proton-deficient nuclei with $Z \leq 12$. This erosion of the shell closure has been related to cross-shell proton-neutron interaction which correlates the $2s_{1/2}$ and $1d_{3/2}$ orbitals with the $1f_{7/2}$ and $2p_{3/2}$ orbitals [15]. Similar cross shell correlations are also responsible for the superdeformed band seen in $^{36}$Ar [16] (see also C. E. Svensson contribution). If cross-shell correlations are indeed the mechanism for the shell erosion, then first indications are already visible in the stable calcium isotopes. Our argument is based on the understanding and explanation of nuclear charge radii, $\langle r_c^2 \rangle$, in the calcium isotopes. The isotopic shifts show a characteristic parabolic shape with a pronounced odd-even staggering when neutrons fill the $f_{7/2}$ orbit and the mass number changes from $A = 40$ to $A = 48$.

The nuclear charge radii can be written as:

$$r_c^2 = r_c^2(\text{mean field}) + r_c^2(\text{correlations}). \quad (4)$$

Mean field calculations, which usually aim at describing nuclear masses, deformation parameters, and radii over a large region of nuclear masses and charges, cannot account for the details of the calcium isotope shifts. The dependence of $\langle r_c^2 \rangle$ on $A$ is usually featureless. Some of the approaches, however, are able to account at least for the near equality of $\langle r_c^2 \rangle$ in $^{40}$Ca and $^{48}$Ca. To this category belong calculations based on the Hartree-Fock method with Skyrme interactions [17], relativistic mean field methods [18], and the extended Thomas-Fermi model with the Strutinsky-integral [19]. Then we will assume that $r_c^2(\text{mean field})$ in equation (4) is constant for all the calcium isotopes and determine the contribution of the correlations using shell-model calculations.

Since here we are interested in the description of calcium isotopes, it is imperative to include states in the vicinity of the $N = Z = 20$ shell boundary. Therefore, the chosen valence space consists of the $d_{3/2}$, $s_{1/2}$, $f_{7/2}$, and $p_{3/2}$ subshells for both protons and neutrons. (Thus $^{28}$Si represents the inert core.) Our calculations [20] reproduce the energies of the intruder states in Sc and Ca, as well as the energies of the low-lying $2^+$ and $3^-$ states (and $B(E2)$) in the even Ca isotopes.

Due to the configuration mixing across the $Z = 20$ shell boundary, protons are lifted from the $sd$ to the $fp$ shell, resulting in a increase of $r_c^2$ that assuming harmonic oscillator wave functions is given by:

$$\delta r_c^2 = \frac{1}{Z} n_{fp}^\pi, \quad (5)$$

where $Z = 20$, b is the oscillator parameter which we assume constant for $A = 40$–48. (Notice that this assumption is supported by the fact that $r_c^2$ is constant in mean field calculations.) $n_{fp}^\pi$ is the number of protons in the $fp$ shell that we obtain from our shell-model calculations.

Figure 5 compares the computed isotopic shifts with data. The trends, parabolic shape and odd-even staggering, are clearly reproduced but the magnitude of the calculated shifts is smaller than the experiment suggests. That could be due to some small dependence of $b$ with $A$ or the fact that we have neglect the $d_{5/2}$ and $f_{5/2}$ orbitals in our model space. However, our results clearly show that cross shell correlations are responsible for the observed isotopic shifts in the calcium isotopes.

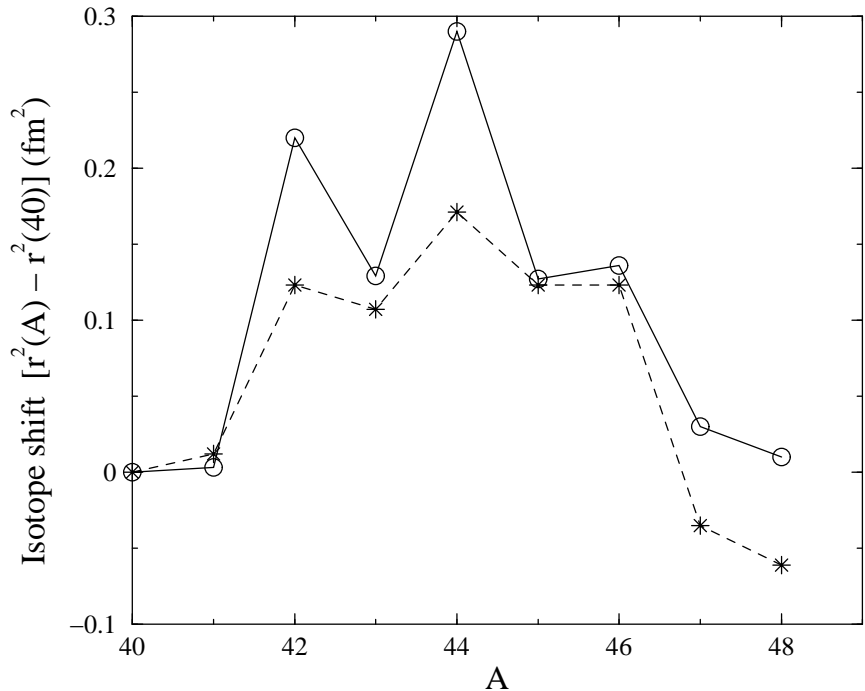

Fig. 5. Isotope shifts in calcium. The experimental data (circles connected by a solid line) and the shell-model results (stars connected by a dashed line) are shown.